\documentclass[twocolumn,prl,aps,superscriptaddress,showpacs]{revtex4}
\usepackage{mathrsfs}
\usepackage{graphicx}
\begin{document}
\title {Dynamical revival of phase coherence in a many-boson system}
\author{Shi-Jie Yang\footnote{Corresponding author: yangshijie@tsinghua.org.cn}}
\author{Sumin Nie}
\affiliation{Department of Physics, Beijing Normal University,
Beijing 100875, China}
\begin{abstract}
We study the quantum dynamics of cold Bose atoms in a double well.
It is shown that self-trapping, as well as population oscillations
are common phenomena associated to nonlinear interactions. For
larger $U/t$, multi-particle tunneling is damped and the quantum
dynamics is dominated by the single-particle tunneling. The
many-body system can be effectively described in a truncated Fock
space. It exhibits coherence-decoherence oscillations in the
temporal evolution. We predict a novel phenomenon of dynamical
revival and collapse of matter wave fields in optical lattices in
regimes near the superfluid-Mott insulator phase boundary.

\end{abstract}
 \maketitle

Tunneling through a barrier is a paradigm of quantum mechanics and
usually takes place on a nanoscopic scale. The existence of
long-range phase coherence, inherent to the description of a
Bose-Einstein condensate (BEC) in terms of a coherent matter wave,
was demonstrated experimentally\cite{Jaksch,Stringari,Greiner1}. In
a simple double well, Josephson oscillations takes place in a
Bose-Einstein condensate (BEC) when the initial population imbalance
is below a critical value\cite{Javanainen,Jack,Zapata}. Above the
critical value and as the repulsive interactions between the atoms
become stronger, a phenomenon of macroscopic quantum self-trapping
where the atoms essentially stay in one well is
predicted\cite{Smerzi,Raghavan} and observed for the experimental
lifetime\cite{Anker,Albiez}.

As to the correlated many-body system, single-particle tunneling and
multi-particle tunneling take effect simultaneously and the
collective motion of the atoms becomes quite complicated. Phase
coherence is very difficult to establish due to the combined
contributions of single- and multi-particle tunneling. Virtually,
self-trapping is not the unique phenomenon of coherent matter waves,
it may happen in a correlated cold atom gas under similar
experimental conditions\cite{Creffield,Milburn}. The physical origin
of self-trapping is, rather than the macroscopic phase coherence of
the system, the nonlinear interactions. Furthermore, population
oscillations between the double wells also take place, given the
interaction is small and the particles are initially imbalanced. The
coherent Josephson oscillations in BECs is just a special case.

In this paper, we explore the dynamical evolution of a correlated
Bose gas in a double well by directly diagonalizing the Hamiltonian
in the Fock space. We attempt to clarify the roles of single- and
multi-particle tunneling in the quantum
dynamics\cite{Winkler,Folling,Zollner,Fischer}. It is shown that the
so-called self-trapping, as well as the population oscillation, are
natural consequences of the dynamical evolution. For larger $U/t$,
the system experiences a coherence-decoherence oscillation. We
further predict a phenomenon of dynamical revival and collapse of
macroscopic matter waves in optical lattices, where the system
parameter is near the superfluid (SF)-Mott insulator (MI) phase
boundary and the state is initially in the deep Mott regimes. This
remarkable phenomenon is caused by the damping of many-particle
tunneling whereas the single-particle tunneling dominating the
quantum dynamics.

We concern with $N$ bosonic atoms confined in symmetrical double
wells (left and right). Making the two-mode approximation and
considering only the lowest energy band, the creation and
annihilation operators ($\hat a_{1,2}^\dagger$ and $\hat a_{1,2}$
respectively) for atoms localized in either side of the well can be
constructed. The Bose-Hubbard Hamiltonian is written
as\cite{Jaksch,Vardi}
\begin{equation}
\hat{H}=-t(\hat a_1^\dagger \hat a_2+\hat a_2^\dagger \hat
a_1)+\frac{U}{2}[\hat n_1(\hat n_1-1)+\hat n_2(\hat n_2-1)],
\label{hamiltonian}
\end{equation}
where $U$ is the repulsion between a pair of bosons occupying the
same site, $t$ is the hopping coefficient, and $\hat n_i$ ($i=1,2$)
are the number operators. Consistence of this model requires
$\frac{1}{2}Un_i(n_i-1)\ll \hbar\widetilde{\omega}_i$, where
$\widetilde{\omega}_i$ is the characteristic frequency of the well
in the harmonic approximation.

The Hamiltonian (\ref{hamiltonian}) is explicitly represented in the
Fock basis set $\{|N,0\rangle,|N-1,1\rangle,\cdots,|0,N\rangle\}$ as
\begin{equation}
H= \left(
          \begin{array}{cccc}
            \frac{U}{2}(N^2-N) & -t\sqrt{N} & \cdots & 0 \\
            -t\sqrt{N} & \frac{U}{2}(N^2-3N+2) & \cdots & 0 \\
            \vdots & \vdots & \ddots & \vdots \\
            0 & 0 & \cdots & \frac{U}{2}(N^2-N) \\
          \end{array}
        \right)
\label{hamiltonian2}
\end{equation}
The general eigenstates are expressed as linear combinations of the
occupation bases, $|\psi_j\rangle=\sum_{k=0}^N c_{jk}|N-k,k\rangle$
($j=0,1,\cdots,N$), which correspond to the eigenvalues $\omega_j$.
The coefficients $c_{jk}$ satisfy the recursive relation
\begin{widetext}
\begin{equation}
-t\sqrt{(N-k)(k+1)}c_{j(k+1)}-t\sqrt{(N-k+1)k}c_{j(k-1)}+[\frac{U}{2}(N^2-2Nk-N+2k^2)-\omega_j]c_{jk}=0.
\end{equation}
\end{widetext}

The temporal evolution of the state is governed by the
Schr\"{o}dinger equation for a given initial state
$|\Psi(0)\rangle$.
\begin{widetext}
\begin{equation}
|\Psi(\tau)\rangle=\sum_{j=0}^N f_j(\tau)|\psi_j\rangle=\sum_{k=0}^N
(\sum_{j=0}^N f_j(0)c_{jk}e^{-i\omega_j \tau})|N-k,k\rangle
\equiv\sum_{k=0}^N g_k(\tau)|N-k,k\rangle, \label{temporal}
\end{equation}
\end{widetext}
with $f_j(0)=\langle\psi_j|\Psi(0)\rangle$ and
$g_k(\tau)=\sum_{j=0}^N f_j(0)c_{jk}e^{-i\omega_j \tau}$.

The population imbalance between the two wells is evaluated by
$Z(\tau)\equiv \langle\Psi |(\hat n_1-\hat
n_2)|\Psi\rangle/N=\sum_{k=0}^N (1-2k/N)|g_k(\tau)|^2$. To depict
the coherence degree of the system, we introduce a characteristic
parameter
\begin{equation}
\alpha(\tau)=\frac{|\lambda_1-\lambda_2|}{\lambda_1+\lambda_2},
\end{equation}
where $\lambda_1$ and $\lambda_2$ are the two eigenvalues of the
single-particle density $\rho_{\mu\nu}(\tau)=\langle\Psi(\tau)|\hat
a_\mu^\dagger \hat a_\nu|\Psi(\tau)\rangle$
($\mu,\nu=1,2$)\cite{Penrose,Mueller}. When $\alpha\rightarrow 1$,
the system is in the coherent (quasicoherent) state since in this
case there is only one large eigenvalue of matrix
$\rho_{\mu\nu}(\tau)$. Accordingly, $\alpha\rightarrow 0$ indicates
the system is in the decoherent or fragmented state because there
are two densely populated natural orbits. One can also use the
quantum fluctuations of the particle number in one of the wells,
$\sigma_1(\tau)=\sqrt{\langle [\hat n_1(\tau)-\langle \hat
n_1(\tau)\rangle]^2\rangle}/ \langle \hat n_1(\tau)\rangle$, to
describe the coherence of the system\cite{Orzel}. In the weak
interaction, strong tunneling limit ($U/t\ll 1$), each atom is in a
coherent superposition of left-well and right-well states. The
ground state of the system is a state with a mean number $N/2$ of
atoms in each well with Poissonian fluctuations
$\sigma_1=\sqrt{N/2}$. The states in each well are quasicoherent
states as the total number of atoms in the system is fixed. In the
opposite limit of strong interactions or weak tunneling ($U/t\gg
1$), the tunneling term is negligible. In this case, the Hamiltonian
is the product of number operators for the left and right wells. The
eigenstates are products of Fock states ($\sigma_1=0$) and are
referred as decoherent states. This regime is analogous to the MI
phase in a lattice system.

In the following we consider two typical cases in which the initial
state is a coherent state or a Fock state. We choose $N=10$ and set
$t=1$ as the units.

{\it Initial coherent states} We first study the quantum dynamics of
an $N$-particle system which is initially in a coherent state,
\begin{equation}
|\Psi(0)\rangle
=(\frac{\hat{a}_{1}^\dagger+\hat{a}_{2}^\dagger}{\sqrt
2})^N|0\rangle.
\end{equation}
The particles are evenly distributed in the two wells, i.e.,
$Z(\tau)\equiv 0$, whereas the number fluctuations persist. Figure 1
shows the temporal evolution of parameters $\alpha$ and $\sigma_1$
for various interactions $U/t$. When $U/t$ is small, $\alpha (\tau)$
fluctuates around unit and the system preserves coherence or
quasicoherence. This is analogous to the SF in optical lattices. As
$U/t$ increases, the value of $\alpha$ lowers, indicating the
coherence is gradually destroyed. However, when $U/t$ further
increases, $\alpha(\tau)$ again exhibits a nearly periodic
oscillation. In the limit of $U/t\gg 1$, the value of $\alpha$
oscillates between unit and zero with a period of $T=\pi/U$,
implying the system experiences revivals and collapses of the
coherence.
\begin{figure}
\begin{center}
\includegraphics*[width=8.5cm]{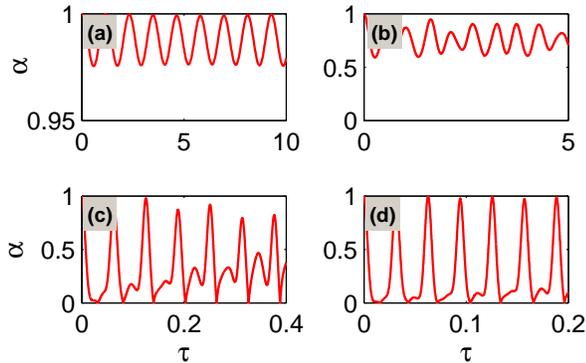}
\caption{(Color online) Temporal evolution of $\alpha(\tau)$ for the
$N=10$ system initiating from the coherent state. From (a) to (d),
$U/t=0.2,2,50,100$, respectively. Periodic revivals of coherence is
clearly enhanced for $U/t\gg 1$.}
\end{center}
\end{figure}

This phenomenon can be explained as follows. The initial state can
be expanded as
$|\Psi(0)\rangle\sim\sum_{k=0}^NC_{N}^{k}(\hat{a}_{1}^\dagger)^{N-k}(\hat{a}_{2}^\dagger)^{k}|0\rangle$.
When $U/t\gg 1$, the Fock states $(a_j^\dagger)^k|0\rangle$ in each
well are the eigenstates and the corresponding eigenenergies are
$Uk(k-1)/2$. The system is then a superposition of products of the
Fock states, which evolves independently as
\begin{widetext}
\begin{eqnarray}
|\Psi(\tau)\rangle&\sim&\sum_{k=0}^NC_{N}^{k}e^{-\frac{i}{2}U(N-k)(N-k-1)\tau}(\hat{a}_{1}^\dagger)^{N-k}e^{-\frac{i}{2}Uk(k-1)\tau}(\hat{a}_{2}^\dagger)^{k}|0\rangle\\\nonumber
&=&e^{-\frac{i}{2}UN(N-1)\tau}\sum_{k=0}^NC_{N}^{k}e^{-\frac{i}{2}U(N-k)k\tau}(\hat{a}_{1}^\dagger)^{N-k}(\hat{a}_{2}^\dagger)^{k}|0\rangle.
\label{revival}
\end{eqnarray}
\end{widetext}
There attaches a time-dependent phase factor in each term. When the
time evolves an integer multiples of $\pi/U$, the system recovers
its initial coherent state. In between this period, superposition
from various terms in (\ref{revival}) cancels and the coherence is
destroyed. Figure 2 displays the coincidence of the full and the
independent evolving dynamics.

\begin{figure}
\begin{center}
\includegraphics*[width=8.5cm]{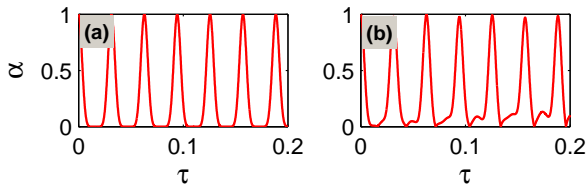}
\caption{(Color online) Comparison of evolution of $\alpha(\tau)$
between (a) independent and (b) full quantum dynamics at $U/t=100$.
The coherence-decoherence oscillation period is $T=\pi/U$. }
\end{center}
\end{figure}

This result has the same gradients of the experiment carried out by
Greiner et al, where they studied the dynamical evolution of the
matter wave field in a 3-dimensional optical lattice\cite{Greiner2}.
While the system is initially prepared in a superfluid state, the
experimental parameters are suddenly changed to the deep Mott
regimes. A periodical collapse and revival of the matter wave fields
were found in the temporal evolution.

{\it Initial Fock states} We now explore our main results for the
many-body system initially in a symmetrical Fock state,
\begin{equation}
|\Psi(0)\rangle=|N/2,N/2\rangle.
\end{equation}
It is found that for all values of $U/t$, $Z(\tau)\equiv 0$, which
implies the atoms are evenly distributed in the two wells. But the
quantum fluctuations of the particle population in each well
persist. Figure 3 shows the coherence degree $\alpha(\tau)$ (solid
lines) and the number fluctuations $\sigma_1(\tau)$ (dot lines)
almost evolve synchronically. Intriguingly, with the enhancement of
interaction, the system gradually exhibits a regular oscillating
feature [Fig.3(c),(d)].

This remarkable phenomenon origins from the damping of the
multi-particle tunneling due to the constraint of energy
conservation. According to the quantum perturbation theory, the
hopping coefficient for single particle tunneling is $t$. For
two-particle co-tunneling it becomes $t^2/U$. As $U\gg t$, the
single-particle tunneling process will dominate the quantum
dynamics. With this consideration, the Hamiltonian can be
effectively approximated in a truncated Fock space by omitting the
multi-particle tunneling,
$\{|N/2-1,N/2+1\rangle,|N/2,N/2\rangle,|N/2+1,N/2-1\rangle\}$, where
only the single-particle tunneling is relevant.
\begin{widetext}
\begin{equation}
\tilde H= \left(
          \begin{array}{ccc}
            U[(N/2)^2-N/2+1] & -t\sqrt{N/2(N/2+1)}  & 0 \\
            -t\sqrt{N/2(N/2+1)} & U[(N/2)^2-N/2] & -t\sqrt{N/2(N/2+1)} \\
            0 & -t\sqrt{N/2(N/2+1)} & U[(N/2)^2-N/2+1] \\
          \end{array}
        \right)
\label{effective}
\end{equation}
\end{widetext}
The broken curves in Fig.3 depict the temporal evolution of
$\alpha(\tau)$ calculated with $\tilde H$ in the truncated Fock
space. The coincidence with the curves in the full Fock space
(Fig.3(c) and (d)) verify our explanation. Further increasing the
interaction will suppress the coherence (Fig.3(d)) as the
single-particle tunneling is also damped. It can be proven from
(\ref{effective}) that in the limit of $U/t\gg 1$, $\alpha(\tau)$
oscillates sinusoidally with a period of $T\sim 2\pi/U$.
\begin{figure}
\begin{center}
\includegraphics*[width=8.5cm]{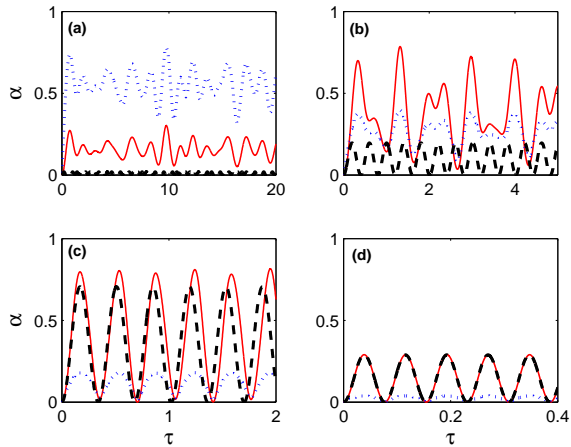}
\caption{(Color online) Quantum dynamics for the $N=10$ system with
the initial Fock state $|\Psi(0)\rangle=|5,5\rangle$. The solid
lines describe $\alpha(\tau)$ and the dotted lines describe the
particle number fluctuations $\sigma_1(\tau)$ in one of the double
well. From (a) to (d), $U/t=0.2,2,10,80$. The broken curves are
$\alpha(\tau)$ calculated in the truncated Fock space.}
\end{center}
\end{figure}

Based on our result in the double well system, we predict a novel
phenomenon of dynamical revivals of macroscopic matter waves for
cold Bose atoms in optical lattices by comparison to the experiment
by Greiner et al\cite{Greiner2}. Suppose the system is initially in
the deep Mott regimes, one suddenly changes the experimental
parameters to regimes near the SF-MI phase boundary, where only the
single-particle tunneling dominates. We can observe creation and
destruction of macroscopic phase coherence in the temporal
evolution. At first sight, our prediction seems the reverse process
of that of Greiners'. The underlying mechanism is quite different.
In our case, macroscopic phase coherence is dynamically generated
from a MI and the single-particle tunneling plays a key role. Most
notably, if the system parameters are directly changed to deep SF
regimes ($U/t\ll 1$), where the multi-particle tunneling becomes as
important as the single-particle tunneling, no macroscopic phase
coherence will occur. Orzel et al\cite{Orzel} had suggested the
possibility of observing the temporal evolution of SF order
parameter. Basically, the bosons are prepared in the number squeezed
Mott state, and then the potential is suddenly reduced into the
superfluid phase. The consequent evolution of the superfluid order
can be deduced from the intensity of interference patterns appearing
when the atoms are released from the trap at sequential times. In a
recent experiment, S. Will et al\cite{Will} observed multi-body
interactions in the time-resolved coherent quantum phase revivals.

Altman et al\cite{Altman} have studied the dynamics of bosons in an
optical lattice using a modified coherent states path integral. They
suggested that a system prepared in the unstable Mott state is
expected to exhibit macroscopic oscillations of the superfluid order
parameter. Our work explicitly clarified the roles of multi- and
single-particle tunneling. We point out that the revival of
coherence takes place only near the phase transition where the
multi-particle tunneling is damped. If the system is directly
changed to deep superfluid regimes, then no revival of the coherence
will be observed. This result is somehow contradict to one's
intuition and is critical for experimental observation.

Finally, we briefly discuss the system evolves from an asymmetrical
Fock state, e.g., $|\Psi(0)\rangle=|10,0\rangle$. Figure 4 shows the
evolution of the population imbalance $Z(\tau)$. In the weak
interaction regimes (Fig.4(a)), $Z(\tau)$ varies between $-1$ to
$1$, indicating a phenomenon analogous to the Josephson oscillation
in BEC. We emphasis the population oscillations is not the Josephson
effect since the system is not in a macroscopic coherent state. As
the interaction increases, $Z(\tau)$ remains positive. Most of the
particles keep staying in one of the wells due to the constraint of
energy conservation. This is the so-called self-trapping phenomenon.
It is notable that in this case there is no revival of phase
coherence.
\begin{figure}
\begin{center}
\includegraphics*[width=7cm]{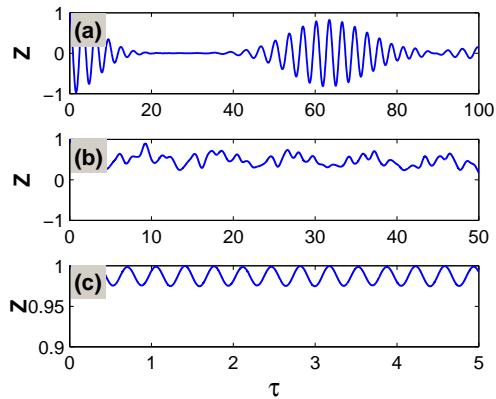}
\caption{(Color online) Temporal evolution of the population
imbalance $Z(\tau)$ for $N=10$ atoms starting from the Fock state
$|10,0\rangle$. From (a) to (c), the interaction $U/t=0.1,0.5,2$.
Self-trapping occurs when $U/t\gtrsim 2$.}
\end{center}
\end{figure}

In summary, we have investigated the dynamical evolution of cold
Bose atoms in a double well. We conclude that self-trapping and
population oscillations between the two wells are consequences of
nonlinear interactions and energy conservation instead of the
macroscopic phase coherence. We predict a novel phenomenon of
revivals and collapses of macroscopic phase coherence from deep MI
regimes in optical lattices. Our results may be helpful to
understand how a macroscopic phase coherence can dynamically be
generated from an incoherent state and the roles of particle
tunneling in the dynamical evolution of quantum many-body system.

We thank M. Ueda and S. Will for helpful discussions. This work is
supported by the National Natural Science Foundation of China under
grant No. 10874018, the 973 Program Project under grant No.
2009CB929101 and "the Fundamental Research Funds for the Central
Universities".

\end{document}